\documentclass[10pt]{article}
\usepackage{amsmath, appendix, authblk, cancel, cleveref, color, empheq, mathtools, mdframed, multirow, sectsty, subfig, txfonts}
\usepackage[top=1.5cm, bottom=1.5cm, left=2.5cm, right=2.5cm]{geometry}

\title{\Large Continuum Model for Pressure Actuated Cellular Structures}
\author{Markus Pagitz and Remco I. Leine\\
 \small markus.pagitz@inm.uni-stuttgart.de \large}
\affil{Institute for Nonlinear Mechanics, University of Stuttgart, 70569 Stuttgart, Germany}
\date{}

\begin{document}
    \maketitle

    \begin{abstract}
        Previous work introduced a lower-dimensional numerical model for the geometric nonlinear simulation and optimization of compliant pressure actuated cellular structures. This model takes into account hinge eccentricities as well as rotational and axial cell side springs. The aim of this article is twofold. \textit{First}, previous work is extended by introducing an associated continuum model. This model is an exact geometric representation of a cellular structure and the basis for the spring stiffnesses and eccentricities of the numerical model. \textit{Second}, the state variables of the continuum and numerical model are linked via discontinuous stress constraints on the one hand and spring stiffness, hinge eccentricities on the other hand. An efficient optimization algorithm that fully couples both sets of variables is presented. The performance of the proposed approach is demonstrated with the help of an examples.
    \end{abstract}

    %-------------------------------------------------------------------------------------------------------

    \section{Introduction}
\begin{figure}[htbp]
            \begin{center}
                \includegraphics[height=0.21\textwidth]{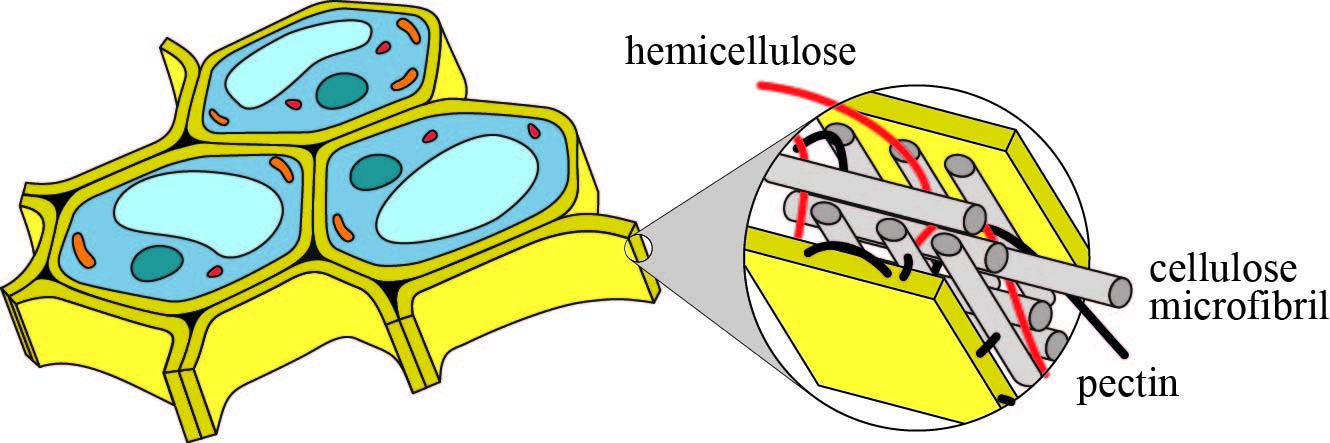}
                \caption{Cell walls are made from a fiber reinforced composite material.}
                \label{pic:Figure_1}
            \end{center}
        \end{figure}

        The multifunctionality and relative simplicity of plant cells is fascinating. Unlike humans and animals, plants do not possess a centralized skeleton and complex control system. Yet they can create their own food through photosynthesis, reproduce, carry considerable external loads and in some cases are even capable of rapid movements. The nastic movement of plants is caused by cell pressure variations of up to 5~MPa \cite{Stahl2006} that require a water flow between neighboring cells \cite{Fleurat-Lessard1997}. Skotheim and Mahadevan~\cite{Skotheim2005} found that the speed of plant movements increases for decreasing cell sizes and pumping distances. Hence it is best if water fluxes occur mainly between neighboring cell layers. Pressurized cells increase the compression strength and thus have a positive impact on a plants overall stiffness. The pressure induced tensile forces are carried by cell walls that are made from a composite material as illustrated in Figure~\ref{pic:Figure_1}.
        Most of the tensile forces are carried by a network of microfibrils that are connected by hemicellulose tethers. This network is embedded in a pectin matrix, a hydrated gel that pushes the microfibrils apart, thus easing their sideways motion \cite{Cosgrove2005}. This allows plants to continuously optimize the position of the microfibrils and thus to alter the mechanical properties. Hence they can minimize stress peaks \cite{Kerstens2001} and maximize the cell wall stiffness \cite{Steudle1977} for given loads. The optimization, design and fabrication of shape changing shell like structures \cite{Pagitz2013} for arbitrary two-dimensional target shapes is hard and most likely an active research area for decades to come. In contrast, although not trivial, the complexity of prismatic cellular structures that can change their shape between one-dimensional target shapes is considerably lower due to their zero Gaussian curvature. Based on previous observations, Pagitz et al~\cite{Pagitz2012} developed a novel concept for adaptive structures that is based on connected rows of pentagonal and hexagonal cells. It is assumed that cell walls are made from an isotropic material and that cell pressures in each row are identical and supplied by an external source. The geometry of each cell can be optimized such that a cellular structure with $n_R$ cell rows deforms into $n_R$ target shapes for given cell row pressures. Compliant pressure actuated cellular structures can undergo large shape changes while being strong and lightweight. Hence their potential application ranges from passenger seats to high-lift devices for aircraft \cite{Pagitz2014}.\\

        The difficulty in realizing this concept lies in the optimization of cell geometries for given target shapes and cell row pressures. Directly optimizing the cell geometries is impractical since it requires a detailed two-dimensional continuum finite element model. It is known from full scale simulations that bending strains are concentrated in regions around cell corners. This observation has been used in \cite{Pagitz2016} for the construction of a lower-dimensional numerical model where each cell side is replaced by axial and rotational springs with hinge eccentricities. The current contribution introduces the associated continuum model for the previously published numerical framework. The side lengths of the numerical model can be optimized for given spring stiffnesses, hinge eccentricities and cell row pressures such that the equilibrium shapes are identical to the target shapes. However, the resulting geometry of the associated continuum model may then be awkward or even impossible to construct. Furthermore, the maximum hinge and cell side stresses in the continuum model may be far from optimal. Hence it is necessary to solve a fully coupled problem such that the structure both meets the target shapes and has a proper geometry. It is shown in \textit{Section~2} that the continuum model can be split into rigid cell corners and elastic cell sides. Furthermore it is shown how the continuum and numerical model are fully coupled. \textit{Section~3} shows that the computation of cell corner geometries is a bilevel optimization problem. The geometry, maximum stresses and spring stiffnesses of elastic cell sides are discussed in \textit{Section~4}. An example is used in \textit{Section~5} to demonstrate the performance of the proposed algorithm. \textit{Section~6} concludes the article.

    %-------------------------------------------------------------------------------------------------------

    \section{Coupling of Numerical and Continuum Model}
\begin{figure}[htbp]
                \begin{center}
                    \subfloat[compliant model]{
                        \includegraphics[height=0.19\textwidth]{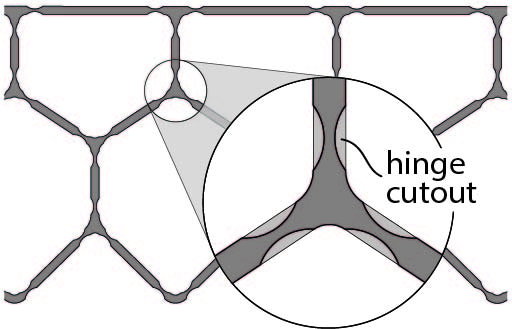}}\hspace{7mm}
                    \subfloat[corners and sides]{
                        \includegraphics[height=0.19\textwidth]{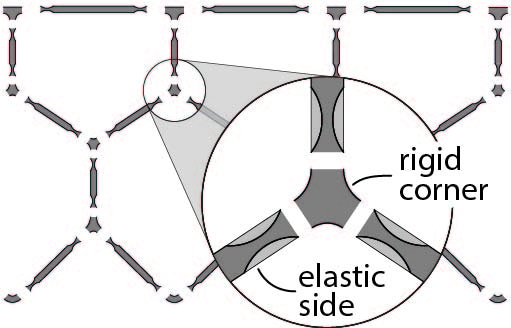}}\hspace{7mm}
                    \subfloat[numerical model]{
                        \includegraphics[height=0.19\textwidth]{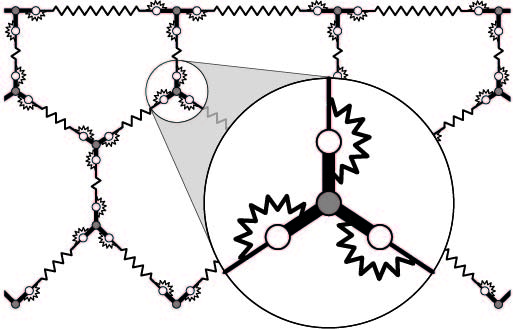}}
                    \caption{Continuum and numerical model for pressure actuated cellular structures.}
                    \label{pic:Figure_2}
                \end{center}
            \end{figure}

        \noindent\textbf{Assumptions}\\
            \noindent Previously published numerical model enables an efficient simulation and optimization of compliant pressure actuated cellular structures for given spring stiffnesses and hinge eccentricities. However, spring stiffnesses and eccentricities are usually not known in advance and might vary during the optimization. The purpose of this article is to introduce a continuum model from which these values can be derived and to couple it to the numerical model. The potential complexity of the continuum model is reduced by making the following assumptions as illustrated in Figure~\ref{pic:Figure_2}:
            \begin{itemize}
                \item continuum model can be split into rigid cell corners and elastic sides
                \vspace{-2mm}
                \item compliant hinges are formed by circular cutouts in rectangular cell sides at both ends
                \vspace{-2mm}
                \item hinges possess a finite bending- and an infinite axial stiffness
                \vspace{-2mm}
                \item central cell sides possess an infinite bending- and a finite axial stiffness.
            \end{itemize}
            \noindent These assumptions support a modular codebase and thus simplify the potential consideration of a wide range of materials and manufacturing processes within the presented framework.
            \vspace{3mm}

        %---------------------------------------------------------------------------------------------------

        \noindent\textbf{State Variables}\\
            The common state variables of the numerical and continuum model are
            \begin{align}
                \mathbf{u}_0 =
                \left[
                \begin{array}{cc}
                    {\mathbf{u}_{\alpha 0}}^\top & {\mathbf{v}_0}^\top
                \end{array}
                \right]^\top
                \hspace{5mm}\textrm{and}\hspace{5mm}
                \mathbf{u}_q =
                \left[
                \begin{array}{ccc}
                    {\mathbf{u}_{\alpha,q}}^\top & {\mathbf{u}_{\kappa,q}}^\top & {\mathbf{v}_q}^\top
                \end{array}
                \right]^\top
            \end{align}
            where the number of reference state variables $\mathbf{u}_0$ is smaller since it is assumed that undeformed cell sides are straight. Additional state variables $\mathbf{w}$ of the numerical model
            \begin{align}
                \mathbf{w} =
                \left[
                \begin{array}{ccc}
                    \mathbf{d}^\top & \mathbf{e}^\top & \mathbf{h}^\top
                \end{array}
                \right]^\top
            \end{align}
            consist of hinge eccentricities $\mathbf{d}$, rotational $\mathbf{e}$ and axial $\mathbf{h}$ springs. These state variables are augmented with the variables $\mathbf{x}$ of the continuum model so that the state variables $\mathbf{w}$ can be expressed as $\mathbf{w}\left(\mathbf{u}_0,\mathbf{x}\right)$.
            \vspace{3mm}

        %---------------------------------------------------------------------------------------------------

        \noindent\textbf{Coupling of Continuum and Numerical Model}\\
            Based on the continuum model, the augmented state variables $\mathbf{x}$ enable the computation of $\mathbf{w}\left(\mathbf{u}_0,\mathbf{x}\right)$ as well as structural stresses $\boldsymbol{\sigma}_q\left(\mathbf{u}_0,\mathbf{u}_q,\mathbf{x}\right)$ of the $q$-th pressure set. The corresponding derivatives are
            \begin{align}
                \mathbf{M}\left(\mathbf{u}_0,\mathbf{x}\right) =
                \left[
                \begin{array}{cc}
                    \mathbf{M}_0 & \mathbf{M}_x
                \end{array}
                \right] =
                \left[
                \begin{array}{cc}
                    \displaystyle \frac{\partial \mathbf{w}}{\partial \mathbf{u}_0} &
                    \displaystyle \frac{\partial \mathbf{w}}{\partial \mathbf{x}}
                \end{array}
                \right]
                \hspace{3mm}\textrm{and}\hspace{3mm}
                \mathbf{N}_q\left(\mathbf{u}_0,\mathbf{u}_q,\mathbf{x}\right) =
                \left[
                \begin{array}{ccc}
                    \mathbf{N}_0 & \mathbf{N}_q & \mathbf{N}_x
                \end{array}
                \right] =
                \left[
                \begin{array}{ccc}
                    \displaystyle \frac{\partial \boldsymbol{\sigma}_q}{\partial \mathbf{u}_0} &
                    \displaystyle \frac{\partial \boldsymbol{\sigma}_q}{\partial \mathbf{u}_q} &
                    \displaystyle \frac{\partial \boldsymbol{\sigma}_q}{\partial \mathbf{x}}
                \end{array}
                \right].
            \end{align}
            In order to couple the continuum and numerical model it is necessary to compute the sensitivities of an equilibrium configuration with respect to state variables $\mathbf{u}_0$ and $\mathbf{w}$. Infinitesimally small variations of current state variables $\mathbf{u}_q$ need to satisfy
            \begin{align}
                \boldsymbol{\Pi}^u_q\left(\mathbf{u}_0+\Delta\mathbf{u}_0,\mathbf{u}_q+\Delta\mathbf{u}_q,\mathbf{w}+\Delta\mathbf{w}\right) = \mathbf{0}.
            \end{align}
            Neglecting higher order terms leads to
            \begin{align}
                \boldsymbol{\Pi}^u_q + \boldsymbol{\Pi}^{u0}_q \Delta\mathbf{u}_0 + \boldsymbol{\Pi}^{uu}_q \Delta\mathbf{u}_q + \boldsymbol{\Pi}^{uw}_q \Delta\mathbf{w} = \mathbf{0}
            \end{align}
            where $\boldsymbol{\Pi}^u_q = \mathbf{0}$ at an equilibrium configuration. The gradient matrix $\mathbf{G}_q$ with respect to state variables $\mathbf{u}_0$ and $\mathbf{w}$ is
            \begin{align}
                \mathbf{G}_q\left(\mathbf{u}_0,\mathbf{w}\right) =
                \left[
                \begin{array}{cc}
                    \mathbf{G}_{0,q} & \mathbf{G}_{w,q}
                \end{array}
                \right] =
                \left[
                \begin{array}{cc}
                    \displaystyle \frac{\partial\mathbf{u}_q}{\partial\mathbf{u}_0} &
                    \displaystyle \frac{\partial\mathbf{u}_q}{\partial\mathbf{w}}
                \end{array}
                \right] =
                -\left(\boldsymbol{\Pi}^{uu}_q\right)^{-1}
                \left[
                \begin{array}{cc}
                    \boldsymbol{\Pi}^{u0}_q & \boldsymbol{\Pi}^{uw}_q
                \end{array}
                \right].
            \end{align}
            Based on the gradients $\mathbf{G}_q$ and $\mathbf{M}$, the matrix $\mathbf{H}$ relates the residual target shape vector $\mathbf{r}_q$ to the state variables $\mathbf{u}_0$ and $\mathbf{x}$ for all $n_R$ pressure sets
            \begin{align}
                \mathbf{H}\left(\mathbf{u}_0,\mathbf{x}\right) =
                \left[
                \begin{array}{cc}
                    \mathbf{H}_0 & \mathbf{H}_x
                \end{array}
                \right] =
                \left[
                \begin{array}{ccc}
                    \left(\mathbf{G}_{0,1} + \mathbf{G}_{w,1} \mathbf{M}_0\right)^\top \mathbf{B}^\top & \ldots & \left(\mathbf{G}_{0,nR} + \mathbf{G}_{w,nR} \mathbf{M}_0\right)^\top \mathbf{B}^\top\\
                    \left(\mathbf{G}_{w,1} \mathbf{M}_x\right)^\top \mathbf{B}^\top & \ldots & \left(\mathbf{G}_{w,nR} \mathbf{M}_x\right)^\top \mathbf{B}^\top
                \end{array}
                \right]^\top
            \end{align}
            where $\mathbf{B}$ is a Boolean matrix. Similarly, the stress gradient matrix $\mathbf{S}_q$ is
            \begin{align}
                \mathbf{S}_q\left(\mathbf{u}_0,\mathbf{x}\right) =
                \left[
                \begin{array}{cc}
                    \mathbf{S}_{0,q} & \mathbf{S}_{x,q}
                \end{array}
                \right] =
                \left[
                \begin{array}{cc}
                    \mathbf{N}_0 +
                    \mathbf{N}_q \left(\mathbf{G}_{0,q} + \mathbf{G}_{w,q} \mathbf{M}_0\right) &
                    \mathbf{N}_x + \mathbf{N}_q \mathbf{G}_{w,q} \mathbf{M}_x
                \end{array}
                \right].
            \end{align}
            It can be seen that the gradient $\mathbf{G}_q$ of the numerical model is used to eliminate the state variables $\mathbf{u}_q$. The $n_\sigma$ independent stress values of all stress vectors $\boldsymbol{\sigma}_q$ can be condensed into a single vector $\boldsymbol{\sigma}$ where the $i$-th entry is
            \begin{align}
                \sigma_i\left(\mathbf{u}_0,\mathbf{u}_q,\mathbf{x}\right) =
                \max_{1 \leq i \leq n\sigma}
                \left[
                \begin{array}{ccc}
                    |\sigma_{1,i}| & \ldots & |\sigma_{nR,i}|
                \end{array}
                \right].
            \end{align}
            Hence, only the pressure set $q$ that leads to a maximum entry $\sigma_i$ is considered. Based on the previously used maximum stress norm, the gradient $\mathbf{S}\left(\mathbf{u}_0,\mathbf{x}\right)$ can be assembled in a similar manner. It should be noted that this gradient is discontinuous since the relevant pressure set for a stress value $\sigma_i$ can vary during the optimization. An objective $F$ that minimizes the difference between current and target values of state variables $\mathbf{u}_0$ and $\mathbf{x}$ is defined as
            \begin{align}
                F\left(\mathbf{u}_0,\mathbf{x}\right) = \frac{1}{2}
                \left[
                \begin{array}{cc}
                    {\mathbf{u}_0-\mathbf{u}_{0,t}}^\top & {\mathbf{x}-\mathbf{x}_t}^\top
                \end{array}
                \right]
                \left[
                \begin{array}{c}
                    \mathbf{u}_0 - \mathbf{u}_t\\
                    \mathbf{x} - \mathbf{x}_t
                \end{array}
                \right]
            \end{align}
            so that the optimization problem can be stated as
            \begin{alignat}{2}
                &\textrm{minimize}                 && F\left(\mathbf{u}_0,\mathbf{x}\right)\\\nonumber
                &\textrm{subject to} \hspace{10mm} && \mathbf{r} = \mathbf{0}\\\nonumber
                &                                  &&\boldsymbol{\sigma} = \boldsymbol{\sigma}_{\textrm{max}}
            \end{alignat}
            where $\sigma_\textrm{max}$ is the maximum allowed stress. This leads to a set of nonlinear equations that are iteratively solved for $\mathbf{u}_0$ and $\mathbf{x}$. In order to avoid the computation of third-order derivatives, target values are dynamically chosen such that $\mathbf{u}_0 = \mathbf{u}_{0,t}$ and $\mathbf{x} = \mathbf{x}_t$ at the onset of each iteration $i$ so that
            \begin{align}
                \left[
                \begin{array}{cccc}
                    \mathbf{I}   & \mathbf{0}   & {\mathbf{H}_0}^\top & {\mathbf{S}_0}^\top\\
                    \mathbf{0}   & \mathbf{I}   & {\mathbf{H}_x}^\top & {\mathbf{S}_x}^\top\\
                    \mathbf{H}_0 & \mathbf{H}_x & \mathbf{0}          & \mathbf{0}\\
                    \mathbf{S}_0 & \mathbf{S}_x & \mathbf{0}          & \mathbf{0}
                \end{array}
                \right]
                \left[
                \begin{array}{c}
                    \mathbf{u}_0^{i+1}-\mathbf{u}_0^i\\
                    \mathbf{x}^{i+1} - \mathbf{x}^i\\
                    \boldsymbol{\lambda}_r^{i+1} - \boldsymbol{\lambda}_r^i\\
                    \boldsymbol{\lambda}_\sigma^{i+1} - \boldsymbol{\lambda}_\sigma^i
                \end{array}
                \right] = -
                \left[
                \begin{array}{c}
                    \mathbf{0}\\
                    \mathbf{0}\\
                    \mathbf{r}\\
                    \boldsymbol{\sigma}-\boldsymbol{\sigma}_{\textrm{max}}
                \end{array}
                \right].
            \end{align}

    %-------------------------------------------------------------------------------------------------------

    \section{Cell Corners}
\begin{figure}[htbp]
                \begin{center}
                    \subfloat[parallel tangents]{
                        \includegraphics[height=0.22\textwidth]{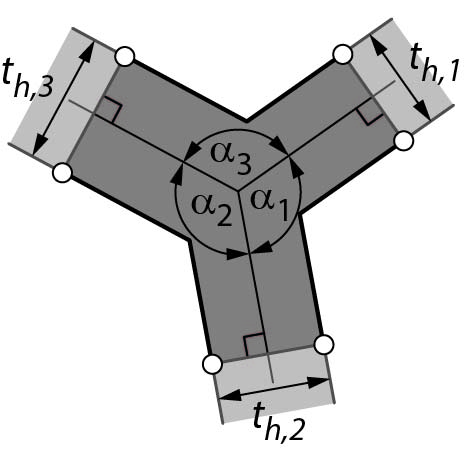}}\hspace{10mm}
                    \subfloat[corner angles]{
                        \includegraphics[height=0.22\textwidth]{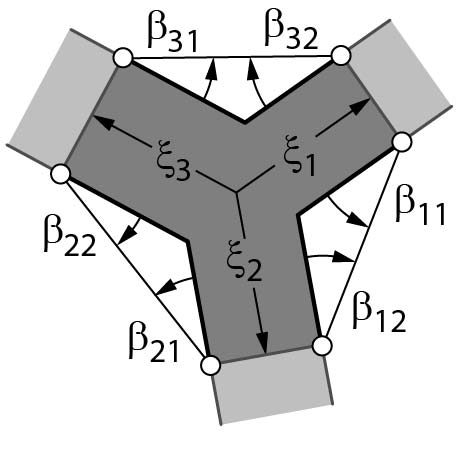}}\hspace{10mm}
                    \subfloat[circular arcs]{
                        \includegraphics[height=0.22\textwidth]{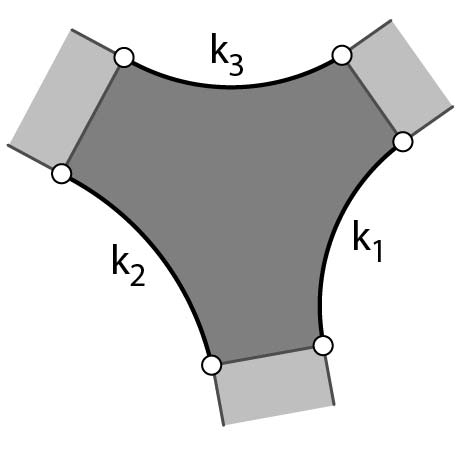}}
                    \caption{Continuum model of a cell corner.}
                    \label{pic:Figure_3}
                \end{center}
            \end{figure}

        \noindent\textbf{Assumptions}\\
            The purpose of cell corners is to continuously connect adjacent cell sides. Their geometry is based on a set of optimality conditions as illustrated in Figure~\ref{pic:Figure_3}. It is assumed that:
            \begin{itemize}
                \item cell corner tangents at the cell side interface are parallel to the central axis
                \vspace{-2mm}
                \item differences between corner angles $\left|\beta_{i2}\left(\boldsymbol{\xi},\boldsymbol{\alpha},\mathbf{t}_h\right) - \beta_{i1}\left(\boldsymbol{\xi},\boldsymbol{\alpha},\mathbf{t}_h\right)\right|$ are minimal
                \vspace{-2mm}
                \item corners are rounded out with circular arcs whose curvatures minimize $\left|k_i\left(\boldsymbol{\xi},\boldsymbol{\alpha},\mathbf{t}_h\right) - k_0\right|$.
            \end{itemize}
            \vspace{3mm}

        %---------------------------------------------------------------------------------------------------

        \noindent\textbf{State Variables}\\
            \noindent State variables $\mathbf{u}_c$ and $\mathbf{x}_c$ of a cell corner are
            \begin{align}
                \mathbf{u}_c\left(\mathbf{u}_0\right) = \boldsymbol{\alpha}\left(\mathbf{u}_0\right)
                \hspace{5mm}\textrm{and}\hspace{5mm}
                \mathbf{x}_c\left(\mathbf{x}\right) = \mathbf{t}_h\left(\mathbf{x}\right).
            \end{align}
            Based on these state variables it is possible to compute the cell corner lengths $\boldsymbol{\xi}\left(\mathbf{u}_c, \mathbf{x}_c\right)$ that, together with $\mathbf{u}_c$ and $\mathbf{x}_c$ define the overall cell corner geometry.
            \vspace{3mm}

        %---------------------------------------------------------------------------------------------------

        \noindent\textbf{Mean Angle Deviations}\\
            \noindent Rounding out a cell corner that connects $n_S$ cell sides with circular arcs requires that
            \begin{align}
                F\left(\boldsymbol{\xi},\mathbf{u}_c,\mathbf{x}_c\right) = \sum_{i=1}^{n_S} \left(\beta_{i2}-\beta_{i1}\right)^2 = 0
            \end{align}
            which can not be satisfied for arbitrary side thicknesses $\mathbf{t}_h$ and cell corner angles $\boldsymbol{\alpha}$. Hence it is generally not possible to avoid kinks between cell corners and sides. Minimizing the mean angle deviations leads to
            \begin{align}
                \tilde{\boldsymbol{\xi}}\left(\xi_s, \mathbf{u}_c, \mathbf{x}_c\right) = \underset{\boldsymbol{\xi}}{\underset{C\left(\boldsymbol{\xi},\xi_s\right) = 0}{\textrm{argmin}}}\ F\left(\boldsymbol{\xi}, \mathbf{u}_c, \mathbf{x}_c\right)
            \end{align}
            where the constraint
            \begin{align}
                C\left(\xi_s, \boldsymbol{\xi}\right) = \xi_s - \sum_{i=1}^{n_S} \xi_i
            \end{align}
            is used to restrict the solution space for $\boldsymbol{\xi}$.
            \vspace{3mm}

        %---------------------------------------------------------------------------------------------------

        \noindent\textbf{Curvature Deviations}\\
            \noindent Circular arcs of a cell corner are required to satisfy
            \begin{align}
                G\left(\tilde{\boldsymbol{\xi}}\left(\xi_s, \mathbf{u}_c, \mathbf{x}_c\right), \mathbf{u}_c, \mathbf{x}_c\right) = \sum_{i=1}^{n_S} \left(k_i-k_0\right)^2 = 0
            \end{align}
            where the target curvature $k_0$ is given and constant throughout a cellular structure. Curvatures $\mathbf{k} = \left[k_1\ \ldots\ k_{nS}\right]^\top$ are
            \begin{align}
                k_i\left(\tilde{\boldsymbol{\xi}} \left(\xi_s,\mathbf{u}_c, \mathbf{x}_c\right),\mathbf{u}_c, \mathbf{x}_c\right) = \frac{2}{L_i} \cos\left(\frac{\alpha_i}{2}\right)
            \end{align}
            where $\mathbf{L} = \left[L_1\ \ldots\ L_{nS}\right]$ are the chord lengths of the circular arcs. The requirement that $G=0$ can not be satisfied for arbitrary cell side thicknesses and cell corner angles.
            \vspace{3mm}

        %---------------------------------------------------------------------------------------------------

        \noindent\textbf{Bilevel Optimization}\\
            Minimizing the curvature deviations leads to a bilevel optimization problem since $G$ depends on optimal hinge eccentricities $\tilde{\boldsymbol{\xi}}$ that are computed by minimizing the mean angle deviations. The optimal sum $\tilde{\xi}_s$ of hinge eccentricities is
            \begin{align}
                \tilde{\xi}_s\left(\mathbf{u}_c, \mathbf{x}_c\right) = \underset{\xi_s}{\textrm{argmin}}\ G\left(\tilde{\boldsymbol{\xi}}\left(\xi_s, \mathbf{u}_c, \mathbf{x}_c\right), \mathbf{u}_c, \mathbf{x}_c\right).
            \end{align}
            The Lagrangian of the first optimization level consists of the objective $F$ and the constraint $C$
            \begin{align}
                \underset{\xi_s,\mathbf{u}_c,\mathbf{x}_c}{\mathcal{L}} \left(\boldsymbol{\xi},\lambda\right) =
                \underset{\mathbf{u}_c,\mathbf{x}_c}{F}\left(\boldsymbol{\xi}\right) +
                \lambda \underset{\xi_s}{C}\left(\boldsymbol{\xi}\right).
            \end{align}
            The variables $\tilde{\boldsymbol{\upsilon}} = \left[\tilde{\boldsymbol{\xi}}\ \ \tilde{\lambda}\right]^\top$ that satisfy the stationarity condition can be iteratively computed by using a Newton based approach. State variables of the $(i+1)-th$ iteration are
            \begin{align}
                \boldsymbol{\upsilon}^{i+1} = \boldsymbol{\upsilon}^i - \mathbf{K}^{-1} \mathbf{f}
            \end{align}
            where the gradient $\mathbf{f}$ and the Hessian $\mathbf{K}$ are
            \begin{align}
                \underset{\xi_s,\mathbf{u}_c,\mathbf{x}_c}{\mathbf{f}}\left(\boldsymbol{\upsilon}\right) =
                \left[
                \begin{array}{c}
                    \displaystyle\frac{\partial\mathcal{L}}{\partial\boldsymbol{\xi}}\\
                    \displaystyle\frac{\partial\mathcal{L}}{\partial\lambda}
                \end{array}
                \right]
                \hspace{5mm}\textrm{and}\hspace{5mm}
                \underset{\xi_s,\mathbf{u}_c,\mathbf{x}_c}{\mathbf{K}}\left(\boldsymbol{\upsilon}\right) =
                \left[
                \begin{array}{cc}
                    \displaystyle\frac{\partial^2\mathcal{L}}{\partial\boldsymbol{\xi}^2} & \displaystyle\frac{\partial^2\mathcal{L}}{\partial\lambda\partial\boldsymbol{\xi}}\\
                    \displaystyle\frac{\partial^2\mathcal{L}}{\partial\boldsymbol{\xi}\partial\lambda} &
                    \mathbf{0}
                \end{array}
                \right] =
                \left[
                \begin{array}{cc}
                    \left(\mathbf{K}^{-1}\right)_{11} & \left(\mathbf{K}^{-1}\right)_{12}\\
                    \left(\mathbf{K}^{-1}\right)_{21} & \left(\mathbf{K}^{-1}\right)_{22}
                \end{array}
                \right]^{-1}.
            \end{align}
            Stationarity of $\tilde{\mathbf{f}} = \mathbf{f}\left(\tilde{\boldsymbol{\upsilon}},\xi_s,\mathbf{u}_c,\mathbf{x}_c\right)$ requires that
            \begin{align}
                \tilde{\mathbf{f}}\left(\tilde{\boldsymbol{\upsilon}} + \Delta \boldsymbol{\upsilon}, \xi_s+\Delta \xi_s, \mathbf{u}_c,\mathbf{x}_c\right) = \mathbf{0}.
            \end{align}
            Linearization of the stationarity condition at the optimum leads to
            \begin{align}
                \tilde{\mathbf{f}}\left(\tilde{\boldsymbol{\upsilon}}, \xi_s,\mathbf{u}_c,\mathbf{x}_c\right) + \frac{\partial \tilde{\mathbf{f}}}{\partial\boldsymbol{\upsilon}} \Delta \boldsymbol{\upsilon} + \frac{\partial\tilde{\mathbf{f}}}{\partial\xi_s} \Delta \xi_s = \mathbf{0}
            \end{align}
            so that
            \begin{align}
                \frac{\partial\tilde{\boldsymbol{\upsilon}}}{\partial\xi_s} =
                -\left(\frac{\partial \tilde{\mathbf{f}}}{\partial\boldsymbol{\upsilon}}\right)^{-1} \frac{\partial\tilde{\mathbf{f}}}{\partial\xi_s}.
            \end{align}
            Hence the sensitivities of $\tilde{\boldsymbol{\xi}}$ with respect to $\xi_s$ are
            \begin{align}
                \frac{\partial \tilde{\boldsymbol{\xi}}}{\partial \xi_s} = \left(\tilde{\mathbf{K}}^{-1}\right)_{12}
                \hspace{5mm}\textrm{and}\hspace{5mm}
                \frac{\partial^2\tilde{\boldsymbol{\xi}}}{\partial {\xi_s}^2} =
                - \left(\tilde{\mathbf{K}}^{-1}\right)_{11}
                \left[
                \begin{array}{ccc}
                    \displaystyle
                    \frac{\partial^3\tilde{\mathcal{L}}}{\partial\xi^2 \partial\xi_1} \left(\tilde{\mathbf{K}}^{-1}\right)_{12}
                    & \ldots &
                    \displaystyle
                    \frac{\partial^3\tilde{\mathcal{L}}}{\partial\xi^2 \partial\xi_{nS}} \left(\tilde{\mathbf{K}}^{-1}\right)_{12}
                \end{array}
                \right]
                \frac{\partial \tilde{\boldsymbol{\xi}}}{\partial \xi_s}.
            \end{align}
            The optimal variable $\tilde{\xi}_s$ of the second level is computed by evaluating
            \begin{align}
                \xi_s^{i+1} = \xi_s^i - \Delta \xi_s
                \hspace{5mm}\textrm{where}\hspace{5mm}
                \Delta \xi_s =
                \begin{cases}
                    \displaystyle
                    \left(\frac{\partial G}{\partial \xi_s}\right)^{-1} G
                    & \textrm{for} \hspace{5mm} \displaystyle
                    \frac{\partial^2 G}{\partial {\xi_s}^2} \leq 0\\
                    \displaystyle
                    \left(\frac{\partial^2 G}{\partial {\xi_s}^2}\right)^{-1}
                    \frac{\partial G}{\partial \xi_s}
                    & \textrm{for} \hspace{5mm}
                    \displaystyle
                    \frac{\partial^2 G}{\partial {\xi_s}^2} > 0.
                \end{cases}
            \end{align}
            The corresponding derivatives are
            \begin{align}
                \frac{\partial G}{\partial \xi_s} =
                \frac{\partial G}{\partial \tilde{\boldsymbol{\xi}}}
                \frac{\partial \tilde{\boldsymbol{\xi}}}{\partial \xi_s}
                \hspace{5mm}\textrm{and}\hspace{5mm}
                \frac{\partial^2 G}{\partial {\xi_s}^2} =
                {\frac{\partial \tilde{\boldsymbol{\xi}}}{\partial \xi_s}}^\top
                \frac{\partial^2 G}{\partial {\tilde{\boldsymbol{\xi}}}^2}
                \frac{\partial \tilde{\boldsymbol{\xi}}}{\partial \xi_s} +
                \frac{\partial G}{\partial \tilde{\boldsymbol{\xi}}}
                \frac{\partial^2 \tilde{\boldsymbol{\xi}}}{\partial {\xi_s}^2}.
            \end{align}
            It can be seen that the first and second level are coupled via the sensitivities $\partial \tilde{\boldsymbol{\xi}}/\partial {\xi_s}$ and $\partial^2 \tilde{\boldsymbol{\xi}}/\partial {\xi_s}^2$. Fully coupling the continuum to the numerical model requires sensitivities of the optimized variable $\tilde{\xi}$ with respect to state variables $\mathbf{u}_c$ and $\mathbf{x}_c$  so that
            \begin{align}
                \frac{d \tilde{\boldsymbol{\xi}}}{d \mathbf{u}_c} =
                \frac{\partial \tilde{\boldsymbol{\xi}}}{\partial\xi_s}
                \frac{\partial \tilde{\xi}_s}{\partial \mathbf{u}_c} +
                \frac{\partial \tilde{\boldsymbol{\xi}}}{\partial \mathbf{u}_c}
                \hspace{5mm}\textrm{and}\hspace{5mm}
                \frac{d \tilde{\boldsymbol{\xi}}}{d \mathbf{x}_c} =
                \frac{\partial \tilde{\boldsymbol{\xi}}}{\partial\xi_s}
                \frac{\partial \tilde{\xi}_s}{\partial \mathbf{x}_c} +
                \frac{\partial \tilde{\boldsymbol{\xi}}}{\partial \mathbf{x}_c}
            \end{align}
            \noindent where
            \begin{align}
                \frac{\partial \tilde{\xi}_s}{\partial \mathbf{u}_c} = -
                \left(\frac{\partial^2 \tilde{G}}{\partial {\xi_s}^2}\right)^{-1}
                \frac{\partial^2 \tilde{G}}{\partial\xi_s\partial \mathbf{u}_c}
                \hspace{5mm}\textrm{and}\hspace{5mm}
                \frac{\partial \tilde{\xi}_s}{\partial\mathbf{x}_c} = -
                \left(\frac{\partial^2 \tilde{G}}{\partial {\xi_s}^2}\right)^{-1}
                \frac{\partial^2 \tilde{G}}{\partial\xi_s\partial\mathbf{x}_c}.
            \end{align}

    %-------------------------------------------------------------------------------------------------------

    \section{Cell Sides}
\begin{figure}[htbp]
                \begin{center}
                    \subfloat[reference configuration]{
                        \includegraphics[height=0.11\textwidth]{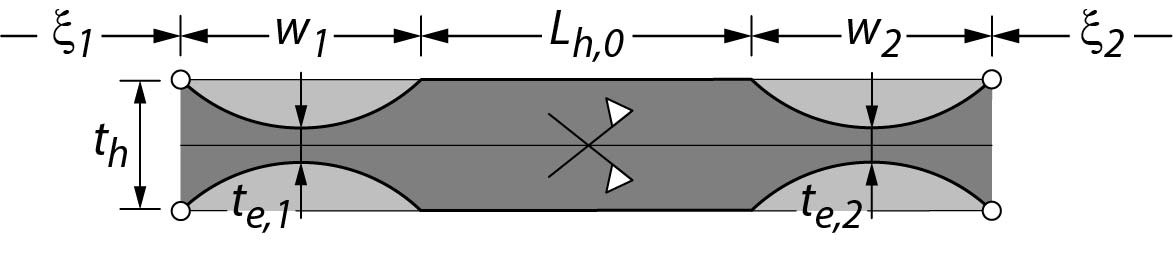}}\hspace{10mm}
                    \subfloat[current configuration]{
                        \includegraphics[height=0.11\textwidth]{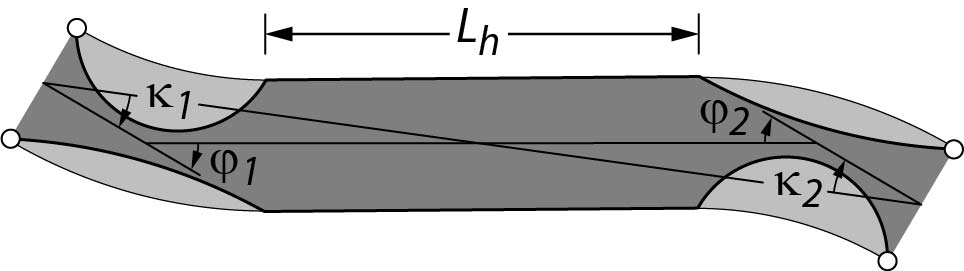}}
                    \caption{Continuum model of a cell side.}
                    \label{pic:Figure_4}
                \end{center}
            \end{figure}

        \noindent\textbf{Assumptions}\\
            Unlike corners, cell sides can undergo elastic deformations. As illustrated in Figure~\ref{pic:Figure_4}, it is additionally assumed that:
            \begin{itemize}
                \item undeformed cell sides possess a reflection symmetry plane
                \vspace{-2mm}
                \item hinge widths depend linearly on hinge thicknesses
                \vspace{-2mm}
                \item deformations and stresses can be described by the euler-bernoulli beam theory
                \vspace{-2mm}
                \item the cell side material is homogeneous, isotropic and in a state of plane strain.
            \end{itemize}
            \vspace{5mm}
            \vspace{3mm}

        %---------------------------------------------------------------------------------------------------

        \noindent\textbf{State Variables}\\
            \noindent State variables $u_{s,0}$, $\mathbf{u}_s$ and $\mathbf{x}_s$ of a cell side are
            \begin{align}
                u_{s,0}\left(\mathbf{u}_0\right) = v_0\left(\mathbf{u_0}\right)
                ,\hspace{5mm}
                \mathbf{u}_s\left(\mathbf{u}\right) =
                \left[
                \begin{array}{ccc}
                    \boldsymbol{\kappa}\left(\mathbf{u}\right) & v\left(\mathbf{u}\right)
                \end{array}
                \right]^\top
                \hspace{5mm}\textrm{and}\hspace{5mm}
                \mathbf{x}_s\left(\mathbf{x}\right) =
                \left[
                \begin{array}{cc}
                    t_h\left(\mathbf{x}\right) & \mathbf{t}_e\left(\mathbf{x}\right)
                \end{array}
                \right]^\top.
            \end{align}
            Unlike cell corners, cell sides can undergo elastic deformations so that both, reference $u_{s,0}$ and current $\mathbf{u}_s$ state variables are considered. Based on these state variables it is possible to compute the hinge widths $\mathbf{w}\left(\mathbf{x}_s\right)$, central cell side lengths $L_{h,0}\left(\mathbf{u}_{s,0},\mathbf{x}_s\right)$, $L_h\left(\mathbf{u}_s,\mathbf{x}_s\right)$ and bending angles $\boldsymbol{\varphi}\left(\mathbf{u}_s,\mathbf{x}_s\right)$ as illustrated in Figure~\ref{pic:Figure_3}. The linear relationship between hinge width $w$ and thickness $t_e$ is
            \begin{align}
                w\left(\mathbf{x}_s\right) = \Xi t_e
            \end{align}
            where the aspect ratio $\Xi$ is constant throughout a cellular structure. The undeformed and deformed central cell side lengths are
            \begin{align}
                L_{h,0}\left(\mathbf{u}_{s,0},\mathbf{x}_s\right) = v_0-\left(\xi_1+\xi_2\right)-\left(w_1+w_2\right)
                \textrm{,}\hspace{5mm}
                L_h\left(\mathbf{u}_s,\mathbf{x}_s\right) = \sqrt{{c_x}^2 + {c_y}^2} - \frac{1}{2}\left(w_1+w_2\right)
            \end{align}
            and the bending angles $\boldsymbol{\varphi}$ are
            \begin{align}
                \boldsymbol{\varphi}\left(\mathbf{u}_s,\mathbf{x}_s\right) =
                \left[
                \begin{array}{c}
                    \kappa_1\\
                    \kappa_2
                \end{array}
                \right]
                +
                \arctan\left(\frac{c_y}{c_x}\right)
                \left[
                \begin{array}{c}
                    1\\
                    1
                \end{array}
                \right]
            \end{align}
            where the vector $\mathbf{c}$ is defined by the hinge centers of the deformed configuration
            \begin{align}
                \mathbf{c}\left(\mathbf{u}_s,\mathbf{x}_s\right) =
                \left[
                \begin{array}{c}
                    v\\
                    0
                \end{array}
                \right]
                +
                \left[
                \begin{array}{r}
                    -\cos\left(\kappa_1\right) \left(\xi_1 + \displaystyle\frac{w_1}{2}\right) - \cos\left(\kappa_2\right) \left(\xi_2 + \displaystyle\frac{w_2}{2}\right)\\
                     \sin\left(\kappa_1\right) \left(\xi_1 + \displaystyle\frac{w_1}{2}\right) + \sin\left(\kappa_2\right) \left(\xi_2 + \displaystyle\frac{w_2}{2}\right)
                \end{array}
                \right].
            \end{align}
            \vspace{3mm}

        %---------------------------------------------------------------------------------------------------

\begin{figure}[htbp]
                \begin{center}
                    \subfloat[]{
                        \includegraphics[height=0.23\textwidth]{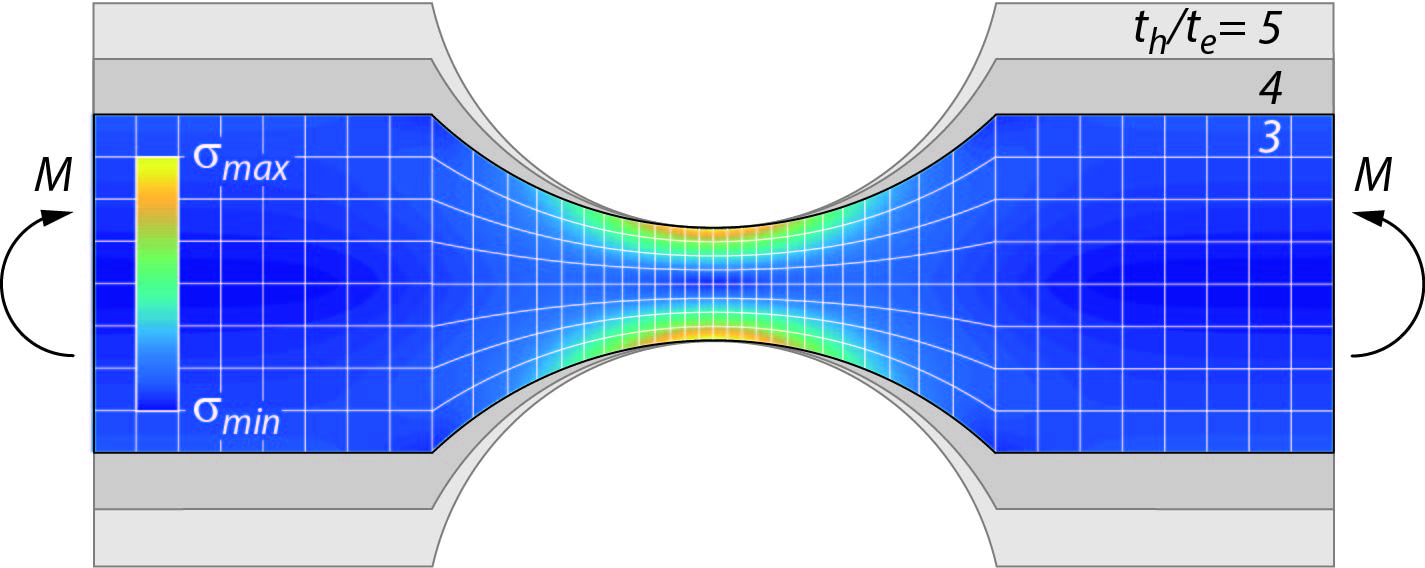}}\hspace{10mm}
                    \subfloat[]{
                        \includegraphics[height=0.23\textwidth]{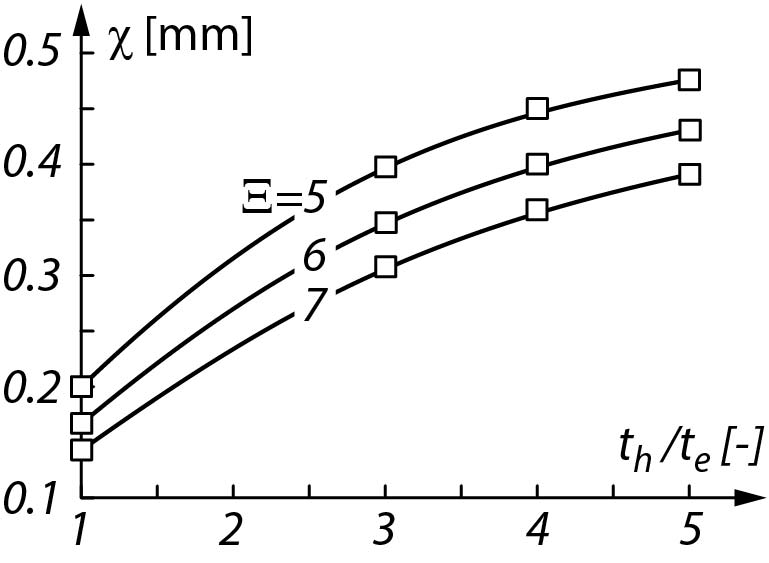}}
                    \caption{Von Mises stresses of a compliant hinge due to a bending moment $M$ for $\Xi = 5$. (b) Interpolated stiffness parameter $\chi\left(\mathbf{x}_s\right)$.}
                    \label{pic:Figure_5}
                \end{center}
            \end{figure}

        \noindent\textbf{Rotational Stiffness}\\
            \noindent The relation between bending moment and hinge rotation is
            \begin{align}
                M_e = e \varphi.
            \end{align}
            The rotational spring stiffnesses $\mathbf{e}\left(\mathbf{x}_s\right)$ of a cell side are fully described by state variables $\mathbf{x}_s$. This is a consequence of the assumption that undeformed cell sides are straight. The rotational stiffness of a single hinge is
            \begin{align}
                e\left(\mathbf{x}_s\right) = \frac{\tilde{E} {t_e}^2}{12} \chi\left(\mathbf{x}_s\right)
            \end{align}
            where the effective Young's modulus $\tilde{E}$ of a material with a Young's modulus $E$ and a Poisson's ratio $\nu$ for a plane strain condition is
            \begin{align}
                \tilde{E} = \frac{E}{1-\nu^2}.
            \end{align}
            The factor $\chi$ is computed with the help of a finite element model (Figure~\ref{pic:Figure_5}) where the discrete points are interpolated with an $\arctan$-function so that
            \begin{align}
                \chi\left(\mathbf{x}_s\right) \approx \frac{1}{\Xi} + \frac{1}{4} \arctan\left(\frac{80-6 \Xi}{100}\left(\frac{t_h}{t_e}-1\right)\right).
            \end{align}
            \vspace{3mm}

        %---------------------------------------------------------------------------------------------------

        \noindent\textbf{Axial Stiffness}\\
            \noindent The relation between axial cell side force and elongation of a central cell side is
            \begin{align}
                F_h = h \left(L_h-L_{h,0}\right)
            \end{align}
            where the axial spring stiffness $h\left(u_{s,0},\mathbf{x}_s\right)$ of the central cell side is
            \begin{align}
                h\left(u_{s,0},\mathbf{x}_s\right) = \frac{\tilde{E} t_h}{L_{h,0}}.
            \end{align}
            The maximum bending moment that acts on a central cell side due to hinge moments $\mathbf{M}_e$ and a differential pressure $\Delta p$ is
            \begin{align}
                M_h =
                \begin{cases}
                    \displaystyle
                    \frac{\Delta p}{2}
                    \left(\frac{L_e}{2} + \frac{M_{e,1} + M_{e,2}}{\Delta p L_e}\right)^2 - M_{e,2} &
                    \mbox{if }\ \left(\Delta p \neq 0\right) \ \mbox{and}\
                    \left(0 < \displaystyle \frac{L_e}{2}+\frac{M_{e,1} + M_{e,2}}{\Delta p L_e} < L_e\right)\\
                    \max\left(|M_{e,1}|,|M_{e,2}|\right) & \mbox{otherwise}
                \end{cases}
            \end{align}
            where the distance between hinge centers is
            \begin{align}
                L_e = L_h + \frac{1}{2}\left(w_1+w_2\right).
            \end{align}
            \vspace{3mm}

        %---------------------------------------------------------------------------------------------------

        \noindent\textbf{Maximum Stresses}\\
            Maximum hinge $\boldsymbol{\sigma}_e\left(\mathbf{u}_{s,0},\mathbf{u}_s,\mathbf{x}_s\right)$ and cell side $\sigma_h\left(\mathbf{u}_{s,0},\mathbf{u}_s,\mathbf{x}_s,\Delta p\right)$ stresses are
            \begin{align}
                \sigma_e = \frac{\rho}{t_e} \left(\frac{6}{t_e}|M_e| + |F_h|\right)
                \hspace{5mm}\textrm{and}\hspace{5mm}
                \sigma_h = \frac{\rho}{t_h} \left(\frac{6}{t_h}|M_h| + |F_h|\right)
            \end{align}
            where the stress reduction factor $\rho$ for the von Mises yield criterion is
            \begin{align}
                \rho = \sqrt{1-\nu+\nu^2}.
            \end{align}

    %-------------------------------------------------------------------------------------------------------

    \section{Example Structures}
\begin{figure}[htbp]
            \centering
            \begin{minipage}[c]{0.23\textwidth}
                \subfloat[]{
                    \includegraphics[width=1\linewidth]{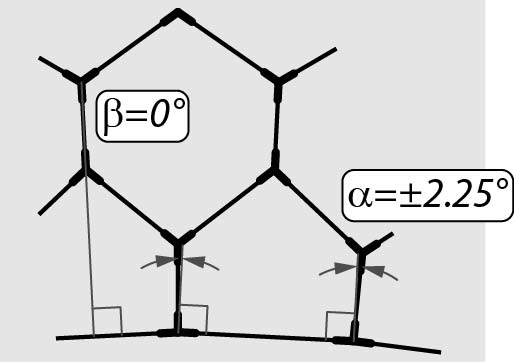}}\vspace{-3mm}

                \subfloat[]{
                    \includegraphics[width=1\linewidth]{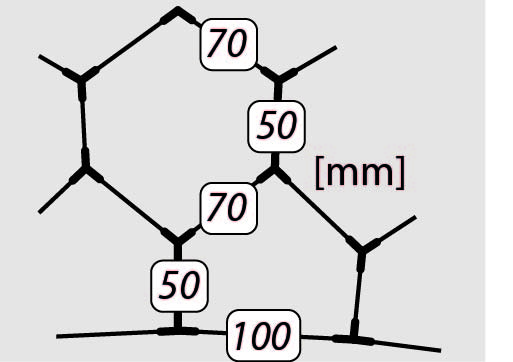}}\vspace{-3mm}

                \subfloat[]{
                    \includegraphics[width=1\linewidth]{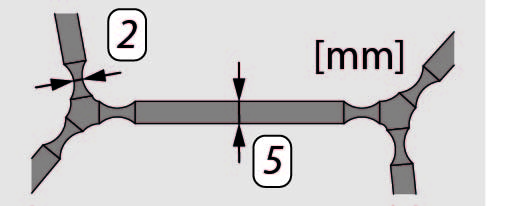}}
            \end{minipage}\hspace{12mm}
            \begin{minipage}[c]{0.6\textwidth}
                \subfloat[]{
                    \includegraphics[width=1\linewidth]{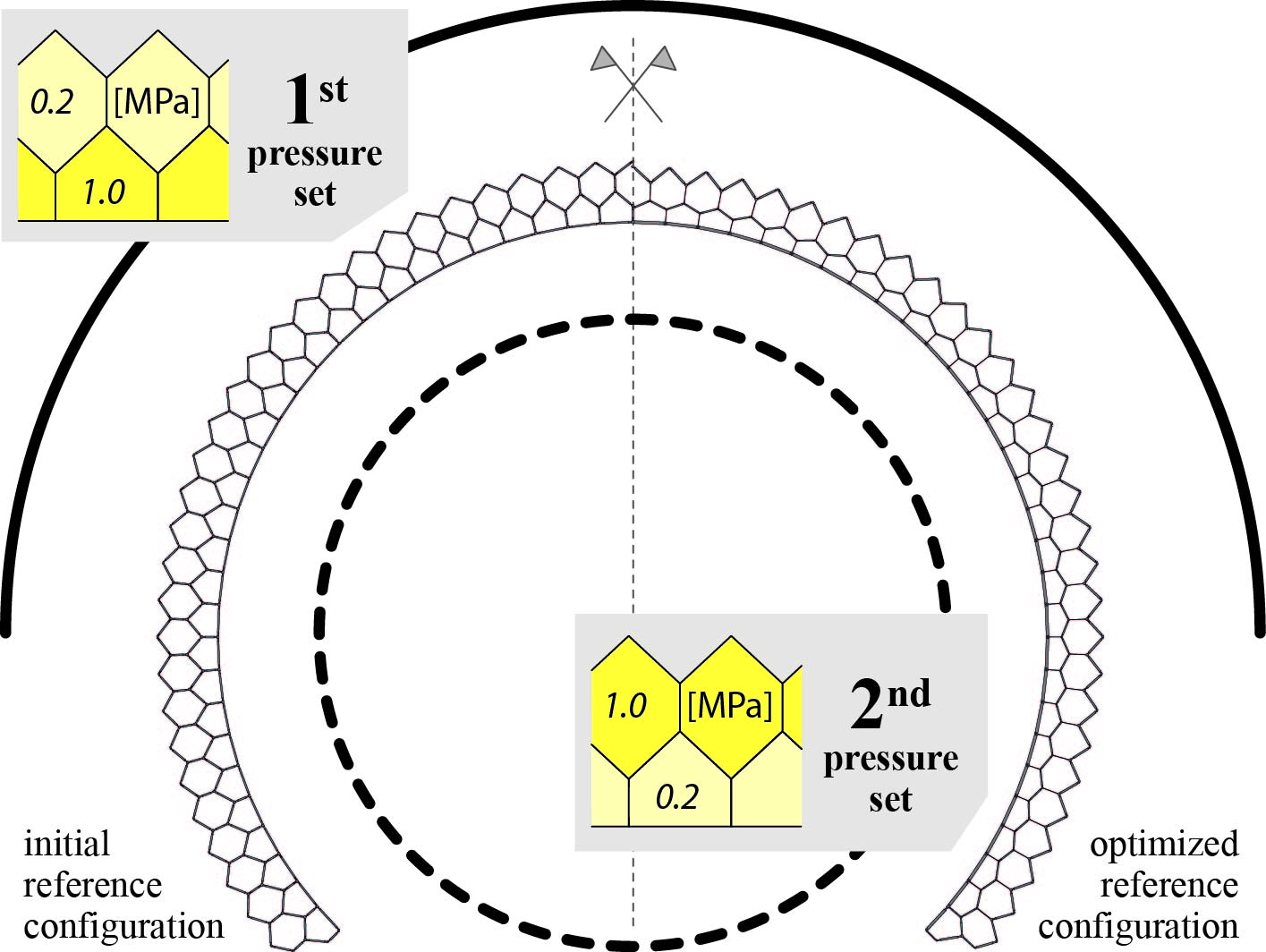}}
            \end{minipage}
            \caption{Reference state variables (a) $\mathbf{u}_{\alpha 0}$, (b) initial cell side lengths $\mathbf{v}_0$ and (c) thicknesses $\mathbf{x}$. (d) Target shapes with associated pressure sets and initial, optimized reference configuration.}
            \label{pic:Figure_6}
        \end{figure}

\begin{figure}[htbp]
            \begin{center}
                \includegraphics[width=1\linewidth]{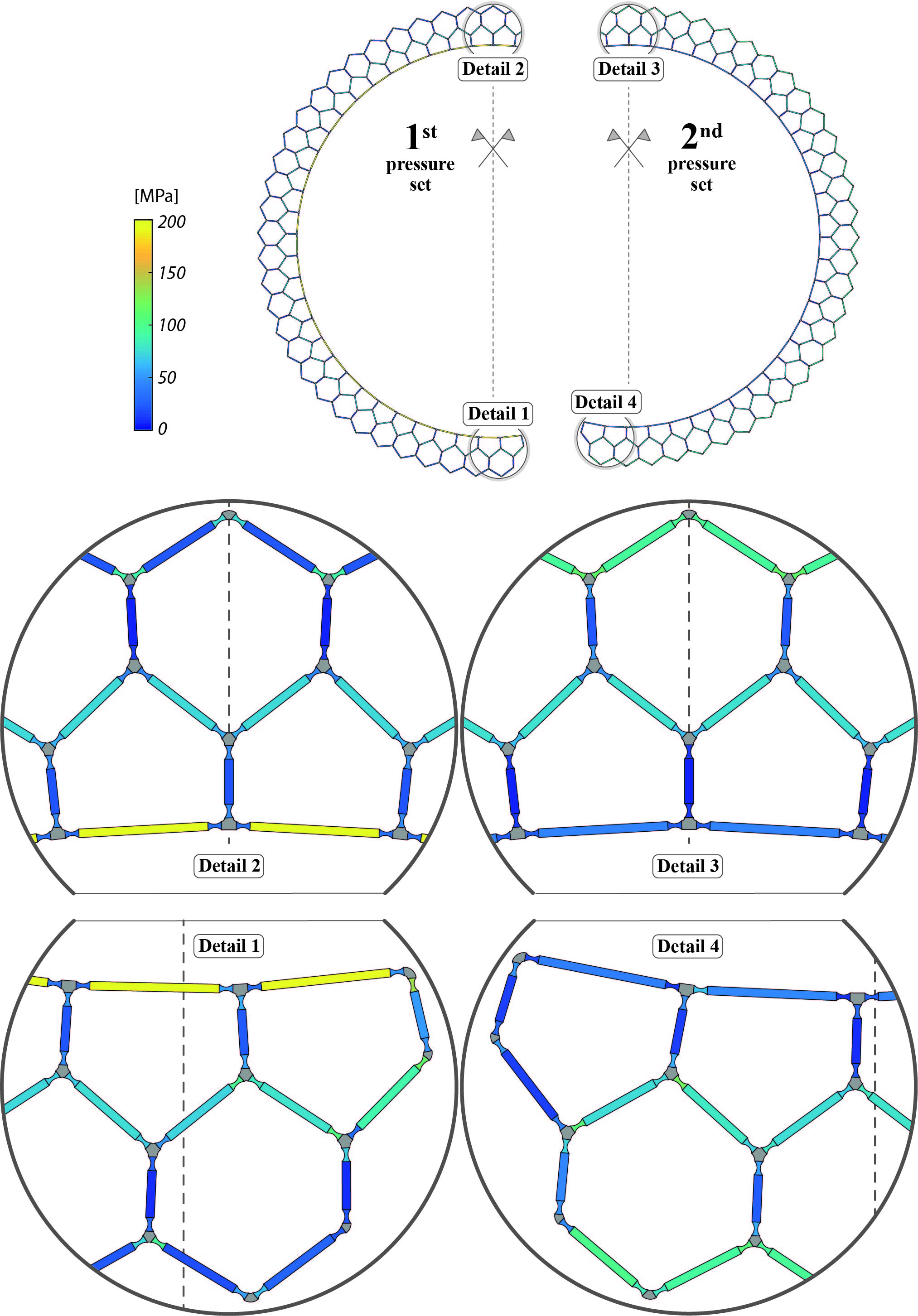}
            \end{center}
            \caption{Equilibrium shapes and hinge, cell side stresses of initial structure.}
            \label{pic:Figure_7}
        \end{figure}

\begin{figure}[htbp]
            \begin{center}
                \includegraphics[width=1\linewidth]{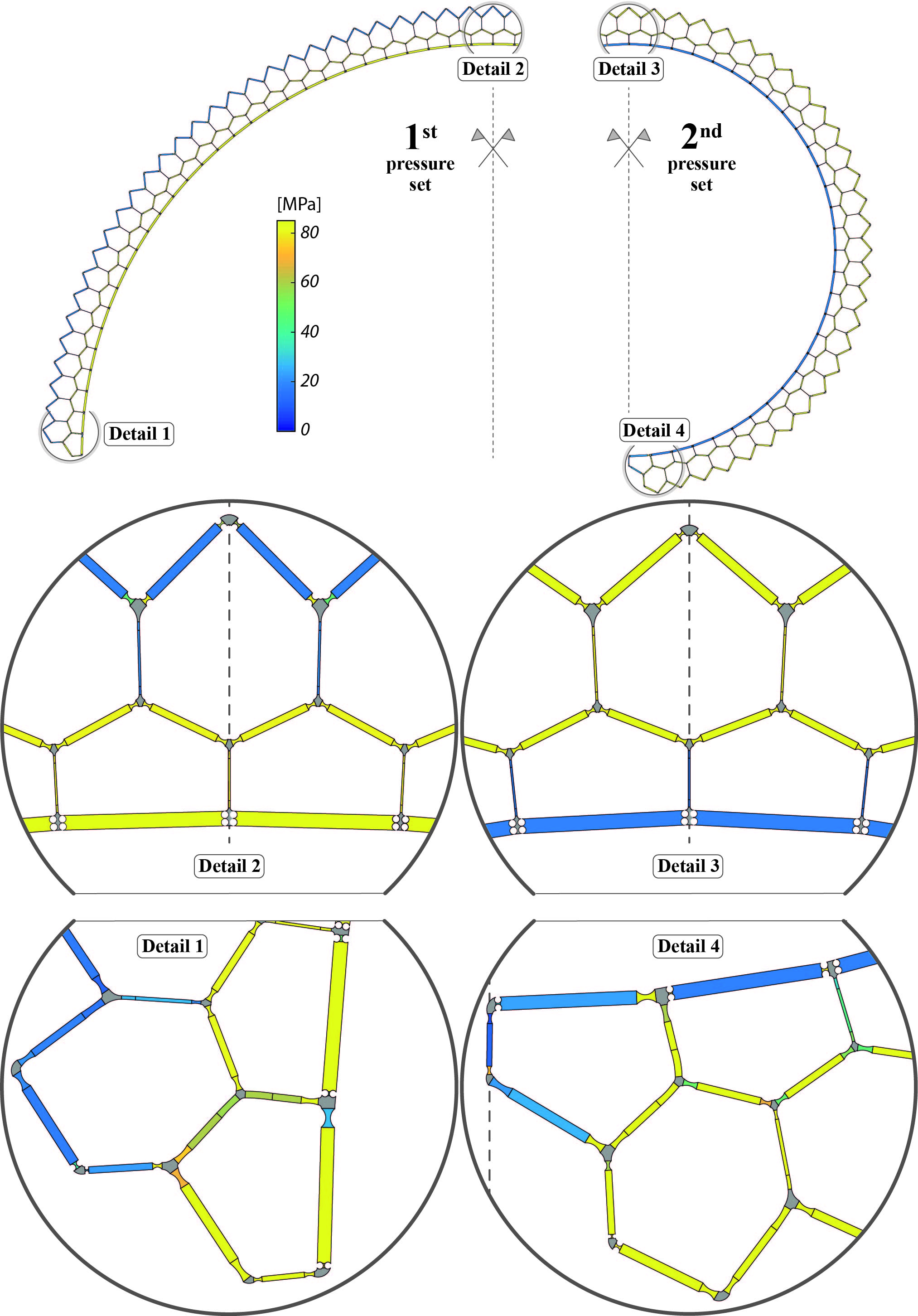}
            \end{center}
            \caption{Equilibrium shapes and hinge, cell side stresses of optimized structure.}
            \label{pic:Figure_8}
        \end{figure}

\begin{figure}[htbp]
            \begin{center}
                \subfloat[]{
                    \includegraphics[height=0.25\linewidth]{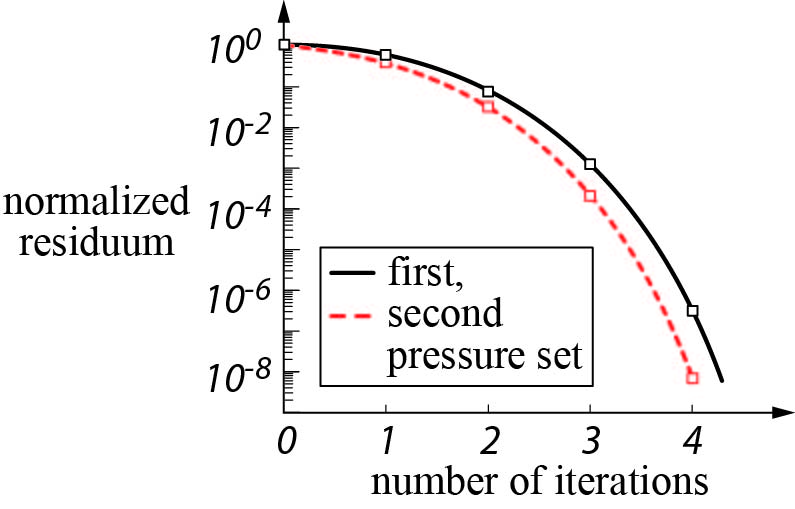}}\hspace{10mm}
                \subfloat[]{
                    \includegraphics[height=0.25\linewidth]{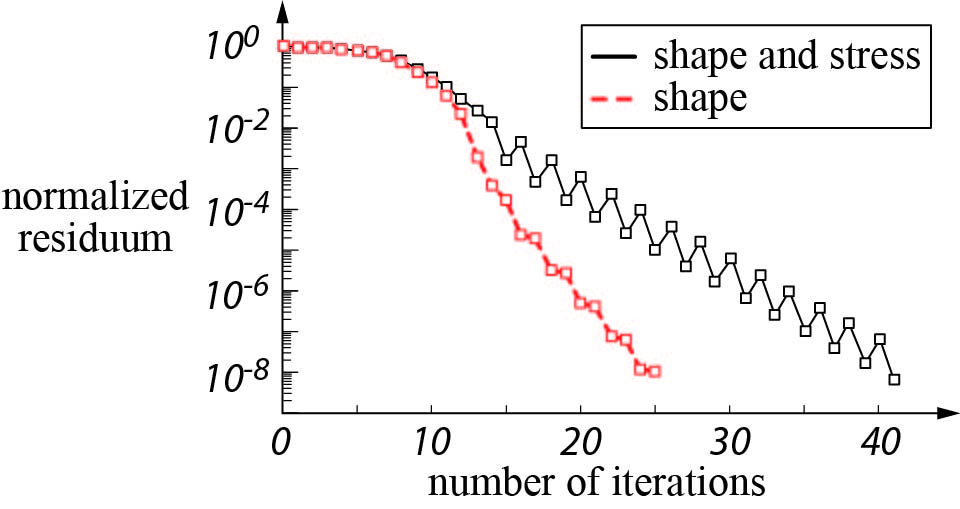}}
            \end{center}
            \caption{Convergence plots for (a) equilibrium shapes of initial structure and (b) optimal cell side lengths and hinge, side thicknesses.}
            \label{pic:Figure_9}
        \end{figure}

        An example structure (Figure~\ref{pic:Figure_6}) that consists of two cell rows with 60 pentagonal and 59 haxagonal cells is used to demonstrate the performance of the proposed algorithm. It is an enhanced version of the example that is used in \cite{Pagitz2016}. The first target shape is a full circle and the second target shape is a half circle. The presented results are based on a complete structural simulation and optimization so that 359 cell side lengths and 1,257 hinge and side thicknesses are optimized. Irrespective of the boundary conditions, edge effects cause varying lengths and thicknesses along the structure so that it is not possible to solve this problem by investigating only a few cells. The used target radius of curvature $r_0$, aspect ratio $\Xi$, effective Young's modulus $\tilde{E}$, stress reduction factor $\rho$ and the maximum stress $\sigma_\textrm{max}$ are
        \begin{align}\nonumber
            r_0 = 2~\textrm{mm}, \hspace{3mm}
            \Xi = 4, \hspace{3mm}
            \tilde{E} = 5,000~\textrm{MPa}, \hspace{3mm}
            \rho = 0.9 \hspace{3mm}\textrm{and}\hspace{3mm}
            \sigma_\textrm{max} = 85~\textrm{MPa}.
        \end{align}
        The equilibrium shapes and maximum hinge, central cell side stresses of the initial (Figure~\ref{pic:Figure_7}) and optimized (Figure~\ref{pic:Figure_8}) structure are shown for both pressure sets. It can be seen that the equilibrium configurations of the optimized structure reassemble a half- and full circle. The optimized continuum model differs significantly from the initial structure and varies between both ends. Furthermore, all hinges and central cell sides are fully stressed for one of the two pressure sets. Convergence plots (Figure~\ref{pic:Figure_9}) show that the equilibrium configurations are computed in four iterations where the maximum step length was limited to $\Delta u_\textrm{max}=3^\circ$ and $\Delta v_\textrm{max}=3~\textrm{mm}$. The optimization requires 41 iterations where the maximum step length was limited to $\Delta v_\textrm{max}=5~\textrm{mm}$ and $\Delta t_\textrm{max}=0.6~\textrm{mm}$. It can be seen that the residuum of the optimization starts to zigzag after 15 iterations. This is due to the fact that the pressure set that causes a maximum hinge or cell side stress can vary during the optimization. Hence, the considered stress constraints are discontinuous. This effect is clearly less pronounced if only the residuum of the target shape is plotted. It can be seen that the structure converges to the desired target shapes in 25 iterations whereas it requires 41 iterations to satisfy the target shapes and stress constraints.

    %-------------------------------------------------------------------------------------------------------

    \section{Conclusions}
        This article introduced a continuum model for compliant pressure actuated cellular structures. It complements previously published numerical model where cell sides are represented by hinge eccentricities and rotational, axial springs. Both models are linked via maximum hinge and side stresses as well as hinge eccentricities and spring stiffnesses. It was shown that the continuum model can be split into rigid cell corners and elastic cell sides. The cell corner geometries were computed with a bilevel optimization approach. The rotational stiffness of a cell side was computed with a detailed finite element model and its axial stiffness with a lower dimensional model. It was demonstrated with the help of an example that the continuum and numerical model can be fully coupled for the shape optimization of compliant pressure actuated cellular structures. An advantage of the presented approach is that the optimized continuum model can be directly send to a rapid prototyping machine. This greatly minimizes the time and costs of future experiments.

    %-------------------------------------------------------------------------------------------------------

\end{document}